\documentclass{article}
\begin{document}
\begin{center}

{\huge {COMMENT}} \\

\bigskip

\bigskip

{\huge {On horizons and the cosmic landscape}}

\bigskip

George F R Ellis\\

Mathematics Department, University of Cape Town, \\
Rondebosch, Cape Town, South Africa.

\bigskip

\end{center}
\begin{abstract}
Susskind claims in his recent book \emph{The Cosmic Landscape}
that evidence for the existence and nature of `pocket universes'
in a multiverse would be available in the detailed nature of the
Cosmic Blackbody Background Radiation that constantly bathes all
parts of our observable universe. I point out that acceptance of
the complex chain of argument involved does not imply possible
experimental verification of multiverses at the present time.
Rather this claim relates only to theoretically possible
observations in the very far future of the universe.
\end{abstract}

A recent book by Susskind \cite{suss05} presents interesting
arguments for the existence of multiverses. Much of that argument
is based on quantum field theory and the non-zero but small value
for the cosmological constant, and is not the concern of this
comment, which focuses on issues to do with the nature of horizons
in cosmology. The point I wish to make is that Susskinds'
contentions about information flows and the Cosmic Background
Radiation (`CBR') in his book do not imply possible astronomical
confirmation of multiverses at any finite time in the history of
the universe.

 The multiverse proposal Susskind espouses states that there exist in a single big
megaverse a vast number of ``pocket universes'' like the expanding
universe domain we see around us, all beyond our observational
reach as they are hidden behind the cosmological horizon. ``Our
cosmic horizon is about fifteen billion light-years away, where
things are moving so rapidly away from us that light from there
can never reach us, nor can any other signal. It is exactly the
same as a black hole horizon - a point of no return. The only
difference is that the cosmic horizon surrounds us, whereas we
surround a black hole horizon. In either case nothing from beyond
the horizon can influence us, or so it was thought. According to
classical physics, those other worlds are forever completely
sealed off from our world''.\footnote{This quote from
\cite{suss05} is taken from the Introduction, available on the web
at
$http://www.twbookmark.com/books/28/0316155799/chapter_{-}excerpt22014.html$
.} The significance is that this apparently means that claims of a
multiverse are not susceptible to observational verification or
disproof; hence their scientific status is open to question:
``Unfortunately the rest of the megaverse of pocket universes is
all in this never-never land beyond the horizon. According to the
classical principles of general relativity, we can wonder all we
want about the existence and reality of these other worlds, but we
can never know. They are metaphysics, not physics''
(\cite{suss05}, p.340; and see also \cite{phil}).

Susskind counters this objection on the basis of what he terms
``the Black Hole War'' - the battle over the fate of information
that falls behind the event horizon of a black hole. He states
that the standard view that all information falling behind the
event horizon is irretrievably lost, has turned out to be wrong.
On the basis of a complex discourse involving the No Quantum Xerox
Principle, Black Hole Complementarity, and The Holographic
Principle, he makes the major claim that is the subject of this
comment:${}^1$ ``The very same arguments that won the Black Hole
War can be adapted to cosmological horizons. \emph{The existence
and details of all the other pocket universes are contained in the
subtle features of the cosmic radiation that constantly bathes all
parts of our observable universe}" (my italics).

Now I do not aim to adjudicate here as to whether the Black Hole War
is won or lost: I simply accept for the sake of the rest of the
discussion that information may not after all be lost when it has
fallen behind the event horizon. The key point is that if this is
so, it does not have the implications set out in the italicized
statement in the previous paragraph, because \emph{the limits on our
present day causal connectivity in cosmology are due to the particle
horizon, not the event horizon} \cite{rin56,pen63,he73,tce}.

While it is true that ``the cosmic event horizon of an eternally
inflating universe is mathematically very similar to the horizon
of a black hole'' (\cite{suss05}, p.340), this has nothing to do
with limits on information available to astronomers today; these
are based on the cosmic particle horizon, which is quite
different. The particle horizon in cosmology (limiting the
`particles' we can see at the present time) is defined in terms of
world lines of matter, and depends on the epoch of observation; it
strictly limits what information about matter we can access
\emph{at the present time}. Furthermore, it exists in any
realistic (almost Friedmann-Lema\^{\i}tre) universe. By contrast,
the event horizon in either cosmology or a black hole context
(limiting the events we will ever be able to see at any time in
our history) is defined by a limiting past null cone, and relates
to the entire world line of the observer, not any particular
observational epoch; whether it exists or not in a cosmological
context is determined by the nature of the end of our history in
the far future \cite{rin56}. An event horizon will not occur in a
cosmology that expands forever with cosmological constant
$\Lambda=0$ in the far future; it exists either at a `Big Crunch'
a finite time from now in the future in a $k=+1$ universe that
recollapses, or at an infinite time in the future in a universe
that expands forever with a non-zero cosmological constant (in
each case there is a spacelike future infinity leading to the
existence of the event horizon\footnote{In the black hole case,
the event horizon is independent of the world line chosen, whereas
in the cosmological case, different event horizons occur for world
lines ending at different points on this spacelike future
infinity.} \cite{pen63,he73,tce}).

The criteria for existence of the two kinds of horizon in a
Robertson-Walker universe with scale factor $a(t)$ are as follows
\cite{rin56}:

- a particle horizon exists if and only if at an arbitrary time
$t_0$ in the universe's history, either the integral
$\int_0^{t_0}dt/a(t)$ converges (if the universe started at time
$t=0$ where $a(t) \rightarrow 0$), or the integral
$\int_{-\infty}^{t_0}dt/a(t)$ converges (if the universe has existed
forever),

- an event horizon exists if and only if at an arbitrary time $t_0$
in the universe's history, the integral $\int_{t_0}^\infty dt/a(t)$
converges (if the universe expands forever in the future) or the
integral $\int^{t_{final}}_{t_0}dt/a(t)$ converges (if the universe
comes to an end at a finite time $t_{final}$ in the future where
$a(t) \rightarrow 0$).

These expressions show how the event horizon relates to the future
(observational events yet to take place), whereas the particle
horizon relates to the past (events that can send signals to be
received by us today).

It is the event horizon that is at the centre of the Black Hole
 information loss paradox; if the resolution of this problem
described by the author is correct, observers would become aware
of any information escaping from their event horizon as they
approached it in the far future of their histories. If information
about what lies beyond a cosmological event horizon were indeed to
emerge, this information would be available to us only at the very
end of the universe. This possibility does not affect present day
observations, which are fundamentally limited not by the event
horizon but by the particle horizon - which excludes all
information from other pocket universes in a supposed megaverse
from reaching us at the present time (in essence: there has not
been enough time since they were formed for light from them to
reach us).

In reality, the limit is even stronger: no significant
cosmological information reaches us at the present time from
beyond the \emph{visual horizon} - defined by the world-lines of
the furthest matter from which we receive electromagnetic
radiation today \cite{ellsto88,ew}.\footnote{This is not related
to the matter moving away from us at the speed of light, as is
often supposed (see \cite{ellroth,lineweaver}).} This matter is
seen by us as the matter that emitted the observed CBR at the time
of decoupling of matter and radiation in the early universe. The
possible far future occurrence of an event horizon in the universe
will not influence CBR data available to us today, \emph{inter
alia} because that information is shaped by the interactions
taking place in the primeval plasma after inflation and before
decoupling, see e.g. \cite{dodel}. Prior information is largely
forgotten during this era - thermalization will destroy any subtle
correlations that may exist at earlier times - and occurrence of
an event horizon in the far future of this epoch is irrelevant.
Possible other `pocket universes' do not enter the calculations
\cite{dodel} of CBR anisotropies.\footnote{Discussions of
`super-horizon modes' and `trans-Planckian effects', see e.g.
\cite{bran}, are all carried out in the context of a single
Robertson-Walker universe domain.}

In summary: The ESA-NASA Planck Surveyor data on CBR anisotropies
will not have coded into it the nature of multiverse regions
enormously more distant from us than a Hubble radius. Attempts to
decipher information about far distant regions of a megaverse
supposedly hidden in that data will not succeed. The black hole
arguments of \cite{suss05} potentially become relevant only at the
end of our destiny when our past light cone merges with the event
horizon, either at future infinity ($t \rightarrow \infty$ on our
world line, with $\Lambda \neq 0$ in that limit), or at the final
crunch in a recollapsing universe ($t \rightarrow t_{final}$ on
our world line, necessarily with $k=+1$); they do not affect
present day astronomy.

Version 2007-03-15.
\end{document}